\documentclass[preprint,onecolumn,nofootinbib]{revtex4}
\usepackage{}
\usepackage{mathrsfs}
\usepackage{graphicx}%Include figure files
\pdfoutput=1
\usepackage[colorlinks=true,linkcolor=blue,urlcolor=blue,filecolor=black,citecolor=red,pdfstartview=FitV,pdftitle={},pdfsubject={},pdfkeywords={},pdfpagemode=None,bookmarksopen=true]{hyperref}
\usepackage{epsfig}
\usepackage{amsmath}
\usepackage{amsfonts}
\usepackage{amssymb}
\usepackage{color}%
\usepackage{dcolumn}% Align table columns on decimal point
\providecommand{\U}[1]{\protect\rule{.1in}{.1in}}

\newcommand{\f}{\begin{equation}}
\newcommand{\ff}{\end{equation}}
\newcommand{\fa}{\begin{eqnarray}}
\newcommand{\ffa}{\end{eqnarray}}

\newcommand{\blue}{\textcolor[rgb]{0.00,0.00,1.00}}

\begin{document}
\title{Characterizing quantum phase transition by teleportation}
\author{Meng-He Wu $^{1,2}$}
\email{menghewu.physik@gmail.com}
\author{Yi Ling $^{2,3}$}
\email{lingy@ihep.ac.cn}
\author{Fu-Wen Shu $^{1}$}
\email{shufuwen@ncu.edu.cn}
\author{Wen-Cong Gan $^{1}$}
\email{ganwencong@gmail.com} \affiliation{$^1$ Center for
Relativistic Astrophysics and High Energy Physics, Department of
Physics, School of Sciences, Nanchang University,
Nanchang 330031, China\\
$^2$ Institute of High Energy Physics, Chinese Academy of Sciences, Beijing 100049, China\\
$^3$ School of Physics, University of Chinese Academy of Sciences,
Beijing 100049, China}
\begin{abstract}
In this paper we provide a novel way to explore the relation
between quantum teleportation and quantum phase transition. We
construct a quantum channel with a mixed state which is made from
one dimensional
quantum Ising chain with infinite length, and then
consider the teleportation with the use of entangled Werner states
as input qubits. The fidelity as a figure of merit to measure how
well the quantum state is transferred is studied numerically. Remarkably we find the first-order derivative of the fidelity with
respect to the parameter in
quantum Ising chain
exhibits a logarithmic divergence at the quantum
critical point. The implications of this phenomenon and possible
applications are also briefly discussed.

\end{abstract}
\maketitle

\section{Introduction}

Quantum phase transition(QPT) is a prominent phenomenon caused by
quantum fluctuations in a many-body system, reflecting the
degeneracy of the ground states\cite{Sachdev:2000}. Unlike thermal
phase transition which is caused by thermal fluctuations and can
always be characterized by some order parameters due to the
symmetry breaking, quantum phase transition is very hard to be
diagnosed when the system is lack of classical order parameters or
manifest symmetry breaking can not be found. In this circumstance
it has been suggested that the entanglement may play a key role in
characterizing quantum phase transition. Therefore, the critical
behavior of some typical quantities which can be used to measure
the degree of entanglement has been extensively investigated in
literature, including the quantum
concurrence\cite{Osborne:2002eia}, entanglement
entropy\cite{Calabrese:2004eu} as well as the fidelity. In
particular, the fidelity as a very crucial notion in quantum
information science, which measures the quality of information
transformation, has been widely used to investigate the occurrence
of quantum phase transition
\cite{Gu:2010fat,Quan:2006dol,Buonsante:2007gfa,Zanardi:2007idg,Zhou:2007GSF,Zanardi:2006GTO,Cozzini:2007GFA}.
Nevertheless, as far as we know, in all previous literature the
fidelity used in this context is the Hilbert-Schmidt fidelity
which is defined as the overlap between two pure quantum states
$F(\lambda ,\lambda+\delta \lambda)=|\langle\varphi
(\lambda)|\varphi (\lambda+\delta \lambda)\rangle|$,
where$|\varphi (\lambda)\rangle$ is a ground state of a many-body
Hamiltonian $\hat{H}(\lambda)$, and $\lambda$ is an external field
parameter. Roughly speaking, in those papers the fidelity just
measure the difference between two ground states when the system
parameter is shifted from $\lambda$ to $\lambda+\delta \lambda$,
while the concept of information loss during the transmission as
described in information science is absent in this context.
Without surprise, in this setup one can find that the fidelity
itself would exhibit extremal behavior
at the critical point
since two ground states
at the critical point are
orthogonal to each other in the thermodynamic limit, which is also
known as the Anderson orthogonality catastrophe\cite{Gu:2010fat,Anderson}.

In this paper we intend to provide a novel way to diagnose the
occurrence of quantum phase transition by quantum teleportation.
In contrast to the strategy as mentioned above, we will construct
a specific quantum channel by picking up two qubits in a one
dimensional
quantum Ising chain with infinite length, and then consider
the fidelity when a specific quantum state is teleported
through this channel. Now, the fidelity becomes a figure of merit\cite{Bose:2007qmt}
to characterize the quality of transmission indeed. Usually the
quantum channel is described by a mixed state with density matrix
$\rho _{c}$ and the fidelity is given by $F\left ( \rho
_{in},\rho _{out} \right)=Tr \left ( \sqrt{\sqrt{\rho _{in}}\rho _{out}\sqrt{\rho _{in}}} \right )$, where $\rho _{in}$ denotes the density matrix of
input mixed state while $\rho _{out}$ corresponds to the density
matrix of output.

Quantum teleportation was originally proposed by C.H. Bennett {\it
et.al.} in 1993\cite{Bennett:1993TAN}. An unknown quantum state
can successfully be transferred through a quantum channel which is
made of a pure but entangled state, given that a classical
information channel could also instruct local observers taking
appropriate operations. Next applying arbitrary mixed state as the
quantum channel associated with the standard teleportation
protocol has been demonstrated in \cite{Bowen:200taa}. Later on a
more specific scheme was proposed to teleport entangled Werner
state\cite{Werner:1989qsw} via thermally entangled states of
two-qubit Heisenberg XX chain in \cite{Yeo:2002TVT}. Inspired by
this scheme we will provide a novel way to construct the quantum
channel with a quantum mixed state which is made from the ground
state of the quantum Ising chain. It is this key point that makes it
plausible to link quantum teleportation to quantum critical
phenomenon in our paper. Thanks to the tensor network techniques
recently developed in
\cite{Roman:2008ITB,Vidal:2005ESO,Verstraete:2008MPS,Orus:2013kga,Vidal:2003EIQ},
we will numerically find the ground states of
quantum Ising chain in terms
of matrix product states(MPS), then construct the quantum channel
by picking up two qubits which could be nearest neighboring or
next-nearest neighboring to each other in the
quantum Ising chain. By tracing out all the other qubits in the chain the
quantum channel will be a mixed state described by a
reduced-density matrix.

Our paper is organized as follows. In next section we will present
the setup for the construction of quantum channel with MPS. Then
in section III we will numerically calculate the entanglement
entropy and the fidelity of the quantum channel when a
Werner state is transferred. More importantly, we will
demonstrate that the first order derivative of the fidelity to the
system parameter will display %an extremal value close to
a logarithmic divergence at the critical point. We
conclude this paper with some discussion on the implications and
possible applications of this phenomenon.

\section{Basic setup}\label{Setup}

\subsection{The ground states of quantum Ising chain in terms of MPS }
In this subsection we will present the setup for the quantum
channel of teleportation. We start with the one-dimensional Ising
model composed of an infinite spin chain, which is one of the
simplest models in many-body physics and exactly solvable
\cite{Pfeuty:1970doi}. The Hamiltonian of the
quantum Ising chain
considered in our paper is given by
\begin{equation}\label{eq:eps}
\hat{H}=\sum_{j=1}^{\infty }\sigma_{1}^{j} \sigma_{1}^{j+1}+\lambda\sum_{j=1}^{\infty }\sigma_{3}^{j}.
\end{equation}
which only involves the neighboring interactions of spins and
$\sigma _{1}=\sigma _{x}$, $\sigma _{3}=\sigma _{z}$ are ordinary Pauli matrices.
%%%%%%%%%%%%%%%%%%%%%%%%%%%%%%

\begin{figure}
\center{
\includegraphics[scale=0.5]{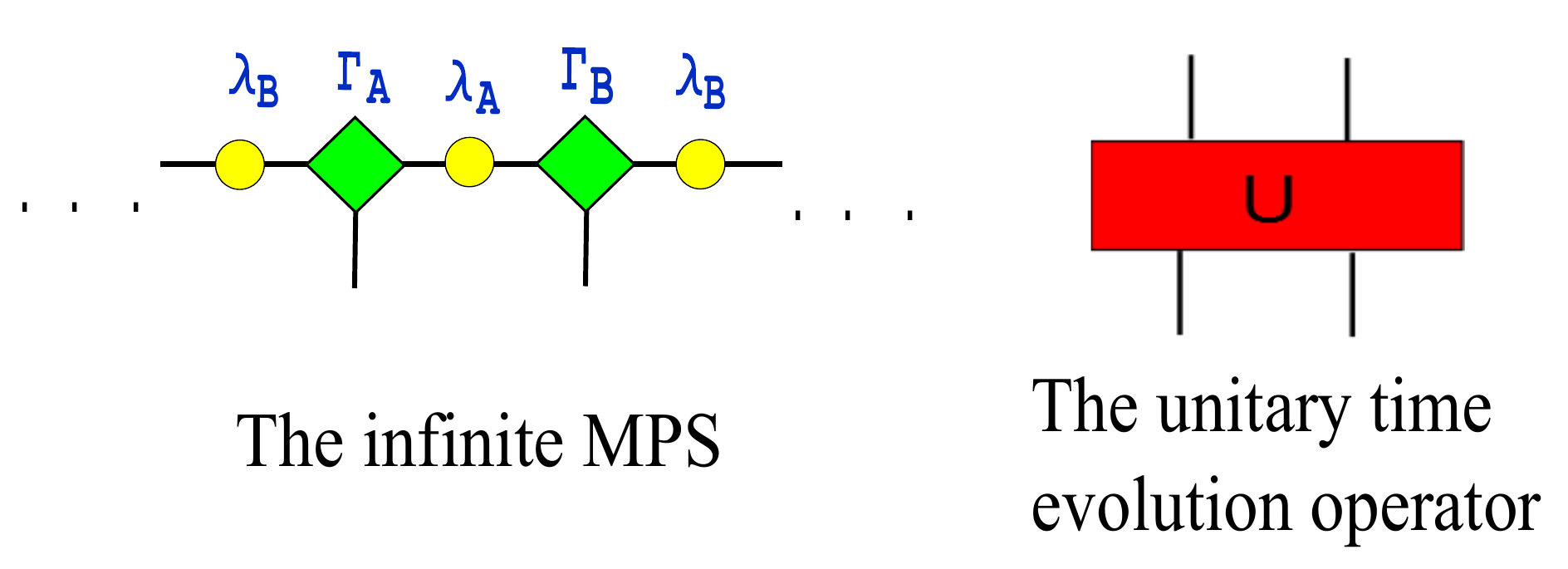} \hspace{0.4cm}

\caption{\label{Fig1} The left is the infinite MPS, and the right is unitary time evolution operator $U=e^{-\hat{H}\delta \tau}$.}}
\end{figure}

%%%%%%%%%%%%%%%%%%%%%%%%%%%%%%
The ground states of above
quantum Ising chain with infinite length
can be
described by matrix product states(MPS) very efficiently
\cite{Verstraete:2006MPS}. For an MPS with infinite qubits, we will
employ infinite time evolving block decimation(iTEBD) algorithm to
simulate the ground states of
quantum Ising chain
\cite{Roman:2008ITB}\cite{Vidal:2007cso}. This algorithm tells us
that starting from any random MPS and performing an imaginary time
evolution by acting the Hamiltonian operators on MPS, one could
finally reach the ground state of the system provided that the
time lasts long enough.

Next we demonstrate the algorithm of iTEBD in our paper briefly,
closely following the logic presented in
\cite{Roman:2008ITB}. First, we construct the MPS with infinite
length. Because the quantum Ising chain in Eq.(\ref{eq:eps}) has
$\mathcal {Z}_{2}$ symmetry, the infinite chain of MPS is only
composed of two distinct pairs of tensors \{$\Gamma_{A}$, $\lambda
_{A}$, {$\Gamma_{B}$, $\lambda _{B}$\} which could be viewed
as the unit cells of the system, where $\lambda _{A}$, $\lambda
_{B}$ are diagonal matrices with non-negative diagonal elements,
as shown in Fig.\ref{Fig1}. Second, we build the unitary time
evolution operator $U=e^{-\hat{H}\delta \tau}$ with the use of the
Hamiltonian in Eq.(\ref{eq:eps}), where $\delta \tau$ is a tiny
time step. Third, we perform the unitary operation by acting
$U$ on the infinite MPS and then contract them into a new tensor
$\Theta$, as illustrated in Fig.\ref{Fig2}. Fourth, singular value
decomposition(SVD) is used to decompose $\Theta$ into
individual tensors $X$ and $Y$, as shown in
Fig.\ref{Fig2}. Lastly, we contract $X$ and $Y$ using matrix
$\lambda_{B}^{-1}$ and obtain updated $\Gamma_{A}$ and
$\Gamma_{B}$, as shown in Fig.\ref{Fig2}. So far, we have
finished the process of updating the unit cells of MPS except for
$\lambda_{B}$. By exchanging $\lambda_{A}$ and
$\lambda_{B}$, we repeatedly perform above process until the
ground state is reached within a precision setting.
%%%%%%%%%%%%%%%%%%%%%%%%%%%%%%
\begin{figure}
\center{
\includegraphics[scale=0.5]{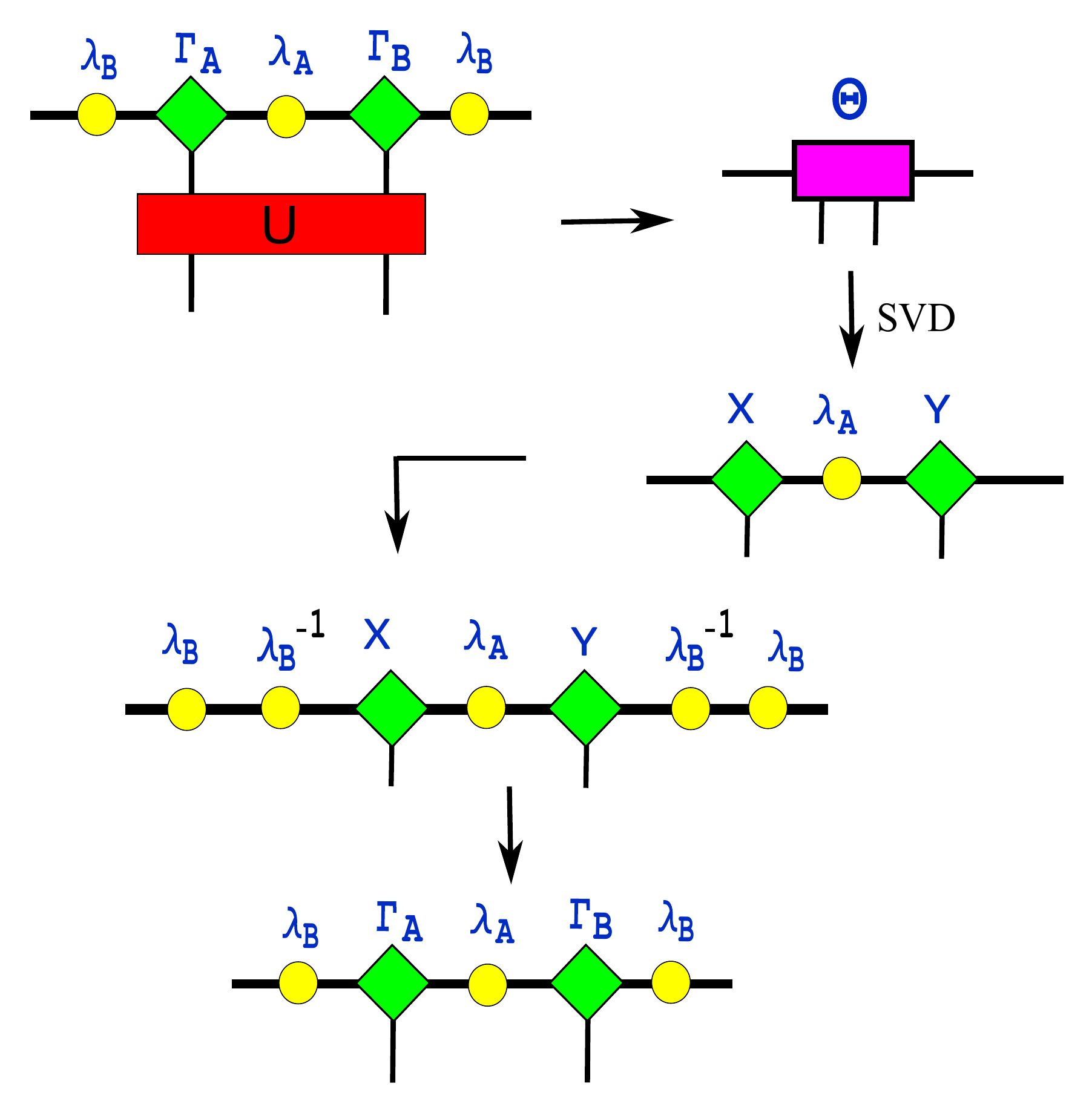} \\ \hspace{0.3cm}
\caption{\label{Fig2}The process of iTEBD algorithm.}}
\end{figure}
%%%%%%%%%%%%%%%%%%%%%%%%%%%%%%
%%%%%%%%%%%%%%%%%%%%%%%%%%%%%%
\begin{figure}
\center{
\includegraphics[scale=0.5]{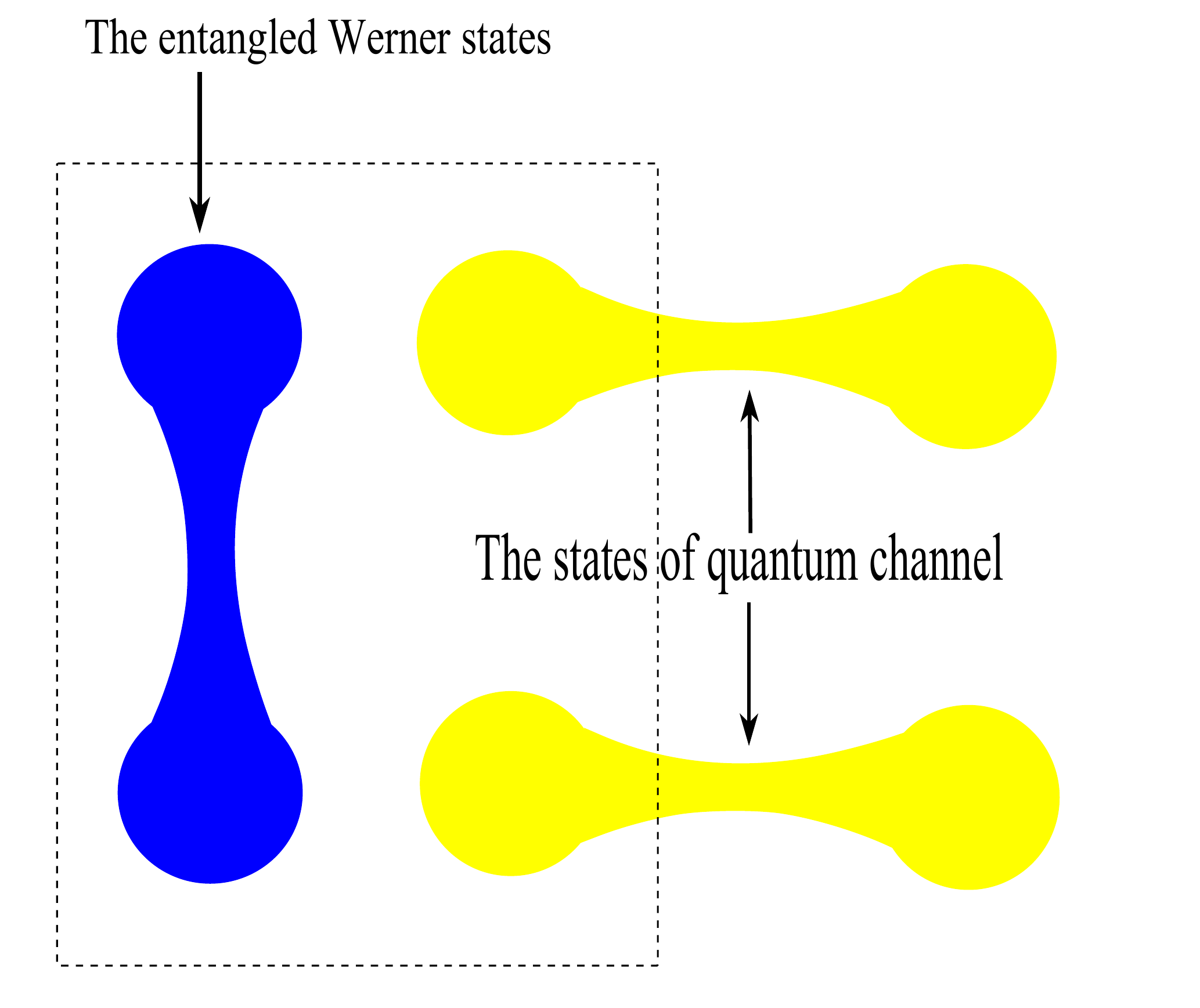} \\ \hspace{0.3cm}
\caption{\label{Fig8} The sketch of the quantum teleportation by taking entangled Werner state as input and using two copies of the mixed states as quantum channel. }}
\end{figure}
%%%%%%%%%%%%%%%%%%%%%%%%%%%%%%

\subsection{Teleportation via mixed entangled states}

In this subsection we will outline our strategy to construct a
quantum channel for teleportation with ground states of
quantum Ising chain. Our main purpose is to teleport a
specific mixed state with a quantum channel which is also made
from a mixed but entangled state. The standard teleportation
protocol has previously been appeared in
\cite{Bowen:200taa,Yeo:2002TVT,Hao:2005ETT} and we will briefly
review their setup as follows. In \cite{Bowen:200taa},
it is originally shown that the standard teleportation with
an arbitrary entangled mixed state $\chi _{AB}$ as quantum channel
is equivalent to a generalized depolarizing channel $\Lambda (\chi
_{AB})$ with probabilities given by the maximally entangled
components of the quantum channel $\chi _{AB}$, i.e.
 \begin{equation}\label{eq:eps4}
\Lambda (\chi _{AB})\rho =\sum_{i}Tr[P_{i}\chi _{AB} ]\sigma _{i}\rho \sigma _{i},
\end{equation}
where $P_{i}=\sigma_{i}P_{0}\sigma _{i}$ ($i=0, 1, 2, 3$) with $P_{0}=|\Phi ^{+}\rangle\langle\Phi  ^{+}|$ and $|\Phi^{+}\rangle=\frac{1}{\sqrt{2}}(|00\rangle+ |11\rangle)$. $\sigma _{0}$ is the
identity matrix and $\sigma _{1}=\sigma _{x}$, $\sigma _{2}=\sigma _{y}$, $\sigma _{3}=\sigma _{z}$ are Pauli matrices. $\rho$ is the single qubit
that we wish to teleport.

Now in this paper, we intend to teleport
entangled Werner states with two qubits $\rho_{W}=\frac{1}{4}(\sigma _{0}\otimes  \sigma
_{0}-\frac{2\gamma+1}{3}\sum_{i=1}^{3}\sigma _{i}\otimes  \sigma _{i})$,
where $0< \gamma \leq 1$. In
paper\cite{Yeo:2002TVT}, thermally entangled states of two-qubit
Heisenberg XX chain are employed to construct the quantum channel.
Given the Hamiltonian of two-qubit Heisenberg XX chain $\hat{H}$,
one can write down the density matrix of the thermal entangled
state as $\rho _{c}=\frac{1}{Z}e^{-\hat{H}/kT}$, where
$Z=Tr(e^{-\hat{H}/kT})$ is the partition function, while $T$ is
the equilibrium temperature and $k$ is Boltzmann constant. Now,
taking entangled Werner state as the input and using two copies of
the above thermal states as quantum channel, see Fig.\ref{Fig8},
the standard teleportation protocol tells us that the density
matrix of the output by teleportation can be written as:
\begin{equation}\label{eq:eps4}
\rho _{out}=\sum_{i,j}Tr[(E^{i}\otimes  E^{j})(\rho_{c}\otimes
\rho_{c})](\sigma _{i}\otimes \sigma _{j})\rho _{in}(\sigma
_{i}\otimes \sigma _{j}),
\end{equation}
where $E^{0}=|\Psi
^{-}\rangle\langle\Psi ^{-}|$, $E^{1}=|\Phi  ^{-}\rangle\langle\Phi  ^{-}|$, $E^{2}=|\Phi
^{+}\rangle\langle\Phi  ^{+}|$, $E^{3}=|\Psi ^{+}\rangle\langle\Psi ^{+}|$, and $|\Phi
^{\pm }\rangle=\frac{1}{\sqrt{2}}(|00\rangle\pm |11\rangle)$, $|\Psi^{\pm
}\rangle=\frac{1}{\sqrt{2}}(|01\rangle\pm |10\rangle)$.
%%%%%%%%%%%%%%%%%%%%%%%%%%%%%%
\begin{figure}
\center{

\includegraphics[width=0.8\textwidth]{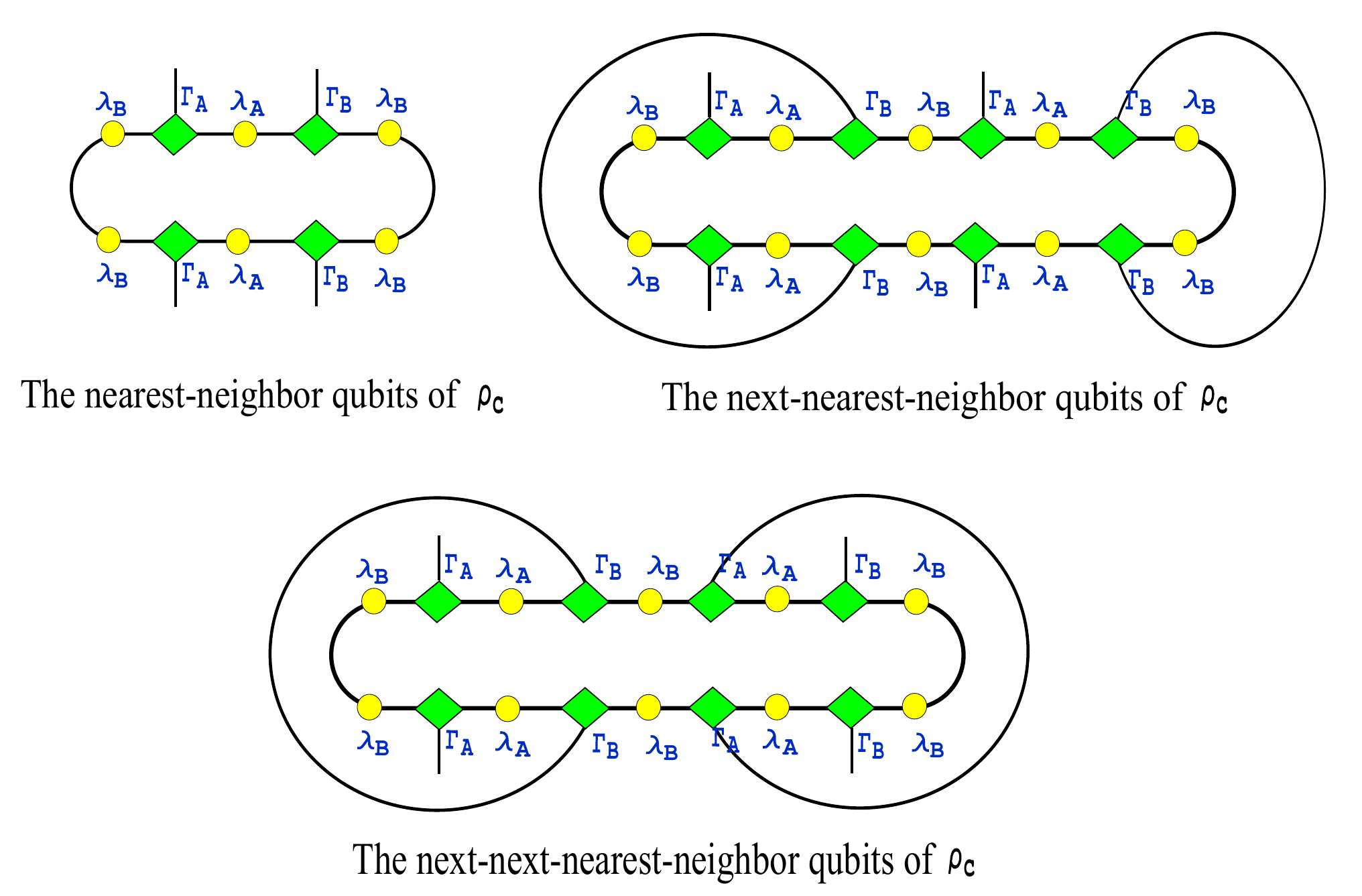} \hspace{0.4cm}
\caption{\label{Fig3}The nearest-neighbor qubits, the next-nearest-neighbor qubits and the next-next-nearest-neighbor qubits of $\rho _{c}$ .}}
\end{figure}
%%%%%%%%%%%%%%%%%%%%%%%%%%%%%%

%%%%%%%%%%%%%%%%%%%%%%%%%%%%%%

\begin{figure}
\center{
\includegraphics[scale=0.34]{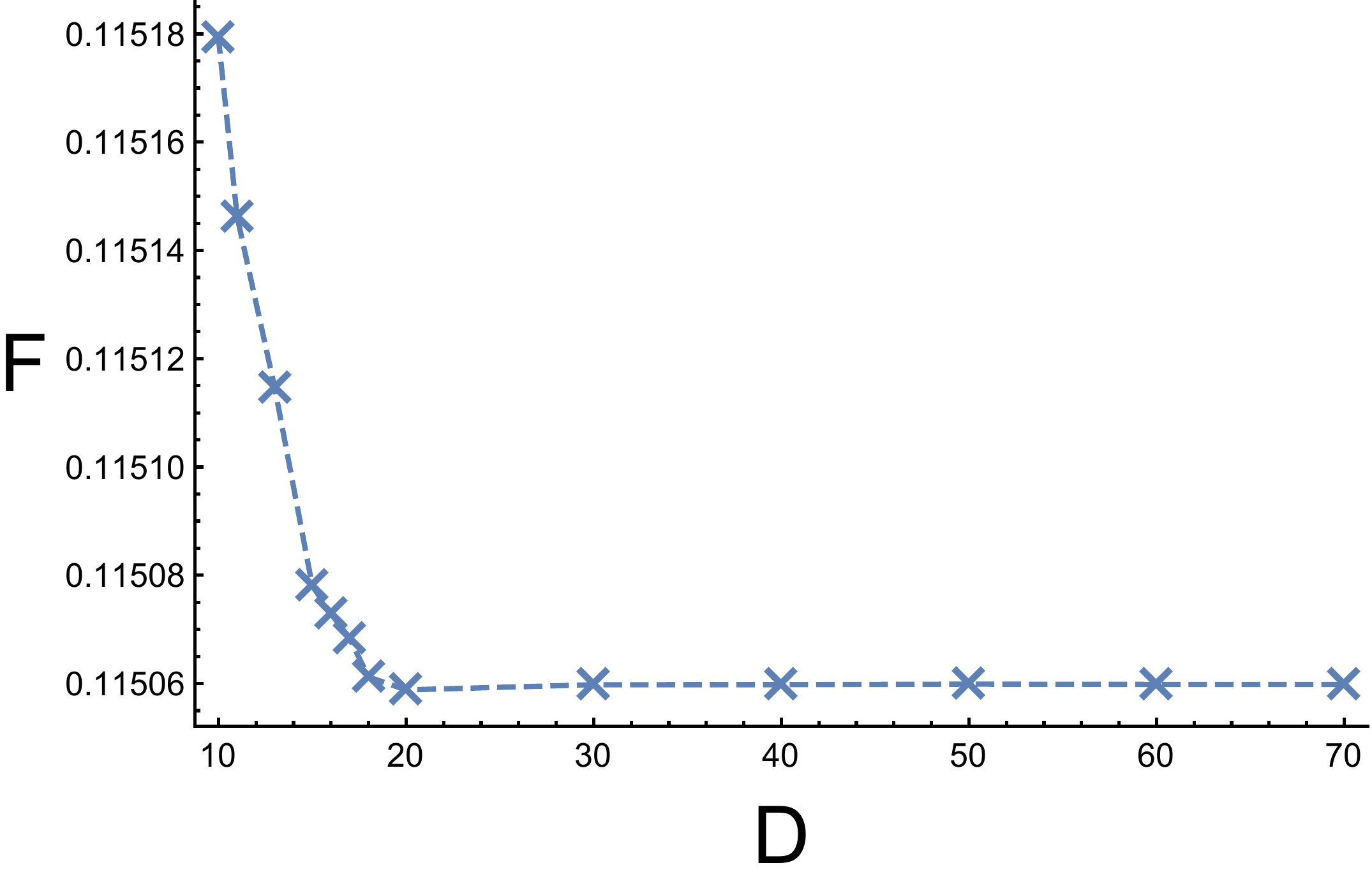} \hspace{0.4cm}
\includegraphics[scale=0.34]{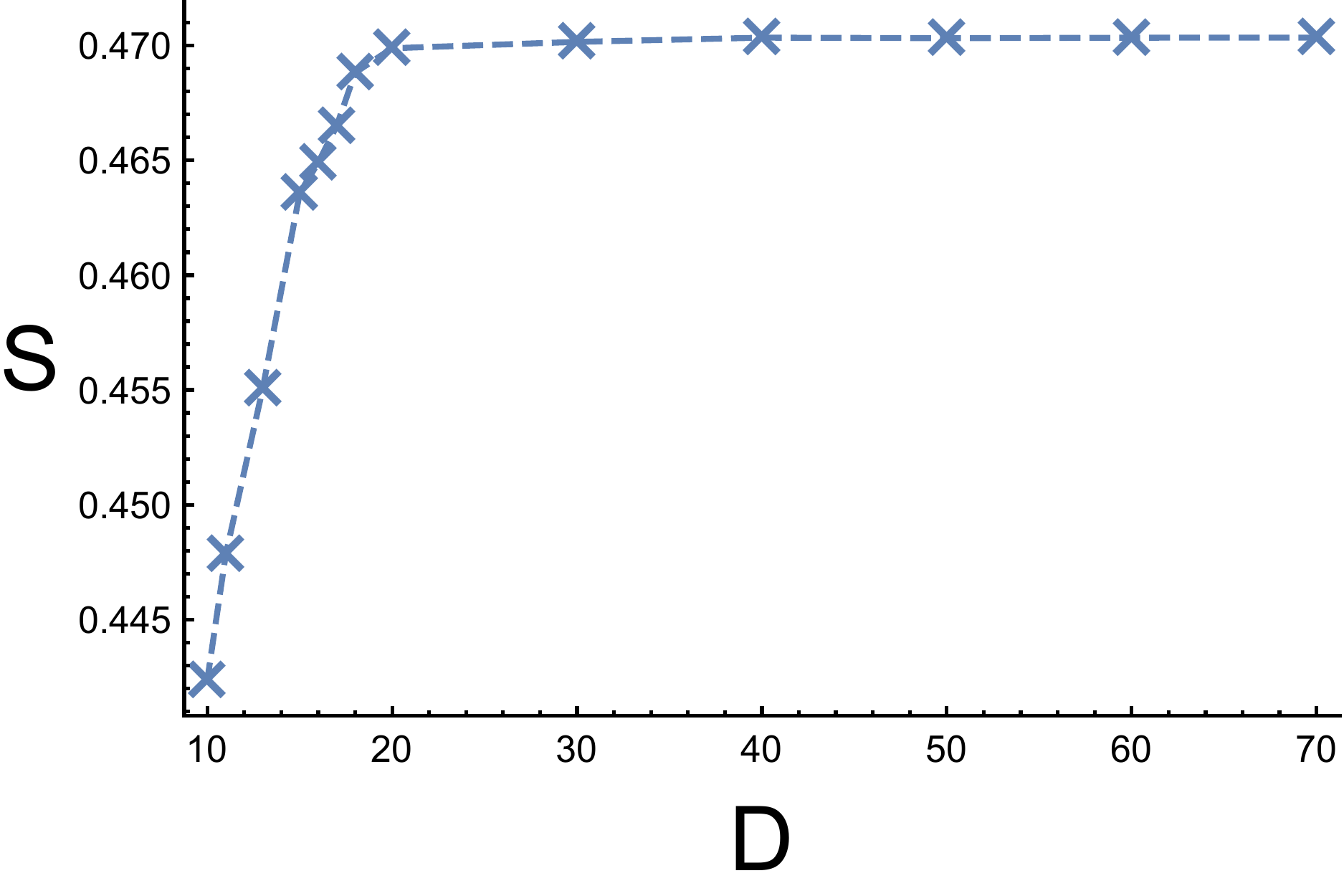} \hspace{0.4cm}
\caption{\label{Figp1} The variations of the fidelity
$F$ and the entropy $S$ with the truncation dimension $D$ at the
critical point for the nearest-neighbor qubits with
$\gamma=1$.}}
\end{figure}

%%%%%%%%%%%%%%%%%%%%%%%%%%%%%%

In our paper one significant change will be considered in order to
investigate the relation between quantum phase transition and
teleportation. Rather than employing a thermal state to construct
the quantum channel, we will apply quantum mixed states of
quantum Ising chain
to construct the channel. Explicitly, given a ground state
of
quantum Ising chain in terms of MPS, we pick up two qubits which could
be the nearest neighboring or next-nearest neighboring or
next-next-nearest neighboring to form the quantum channel. Then by
tracing out all other qubits of the density matrix, a reduced
density matrix of this quantum mixed state $\rho _{c}$ could be
obtained for the quantum channel. We illustrate this process in
Fig.\ref{Fig3}. Similarly, when we use entangled Werner state as
input state $\rho _{in}$ to teleport through this channel, an
output state $\rho _{out}$ with the same expression as in
Eq.(\ref{eq:eps4}) can be obtained.
%%%%%%%%%%%%%%%%%%%%%%%%%%%%%%
\begin{figure}
\center{
\includegraphics[scale=0.4]{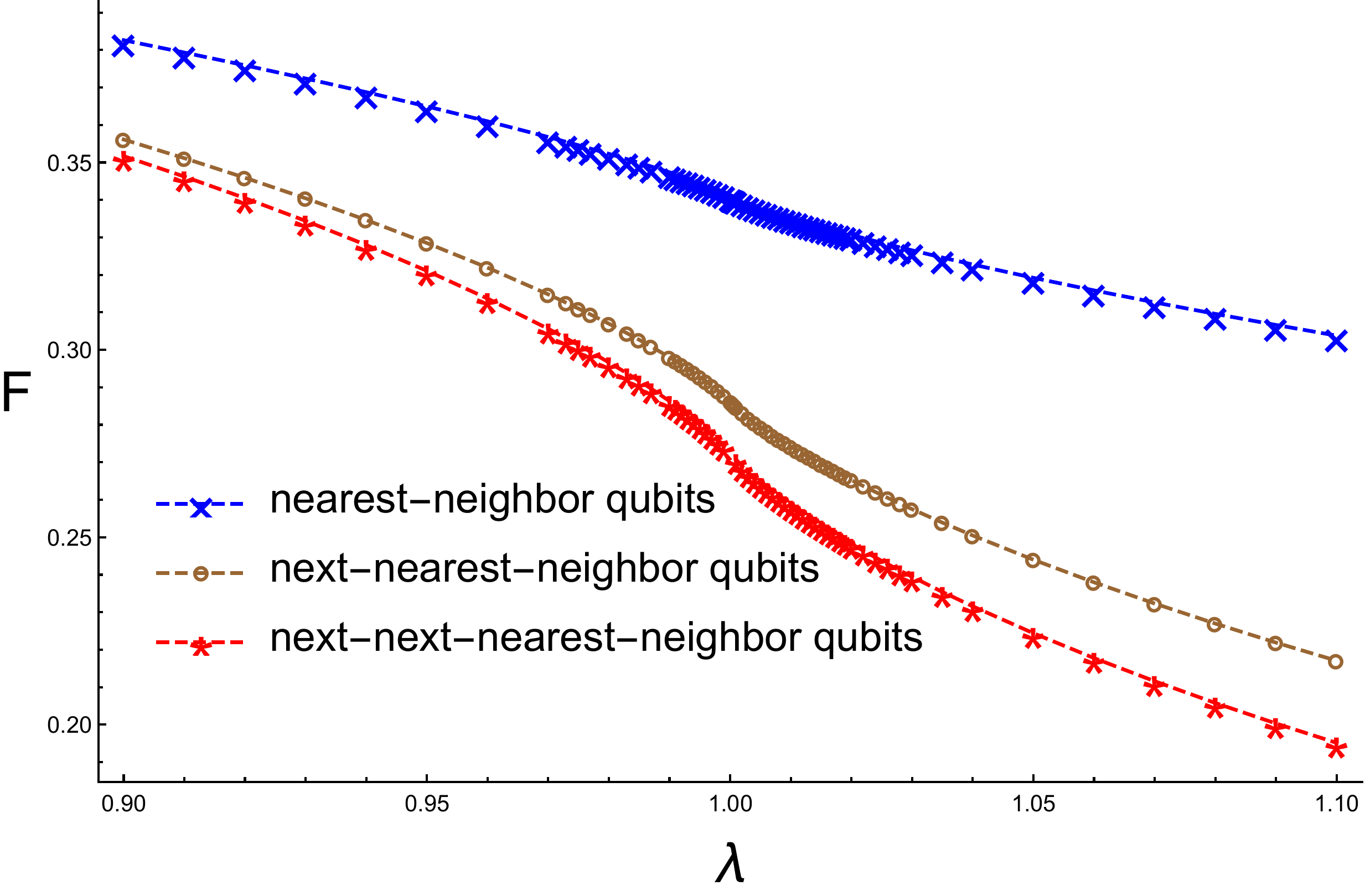} \\ \hspace{0.1cm}
\caption{\label{Fig4}The fidelity as functions of $\lambda$ for the nearest-neighbor
qubits, the next-nearest-neighbor qubits and the
next-next-nearest-neighbor qubits with $\gamma=1$.}}
\end{figure}
%%%%%%%%%%%%%%%%%%%%%%%%%%%%%%
The fidelity of this system now can be evaluated by the difference
between $\rho _{in}$ and $\rho _{out}$, which is given by:
\begin{equation}\label{eq:eps5}
F\left ( \rho
_{in},\rho _{out} \right)=Tr \left ( \sqrt{\sqrt{\rho _{in}}\rho _{out}\sqrt{\rho _{in}}} \right).
\end{equation}
In next section we will report our numerical results about the
fidelity with the variation of the system parameter $\lambda$, and
disclose its critical behavior at the critical point of phase
transition.

\section{The fidelity at quantum critical point }
In this section we present the numerical results of fidelity with
a focus on its behavior
at the critical point of
quantum Ising chain,
which is located at $\lambda=1$. In general, the fidelity $F\left
( \rho _{in},\rho _{out} \right)$ depends on parameters $\lambda$
and $\gamma$. As an example, in Fig.\ref{Fig4} we demonstrate the
fidelity as a function of $\lambda$ for the nearest-neighbor
qubits, the next-nearest-neighbor qubits and the
next-next-nearest-neighbor qubits with $\gamma=1$, respectively.

From Fig.\ref{Figp1}, we obviously observe that the
fidelity $F$ and the entropy $S$ are convergent when the
the truncation
dimension of MPS $D$ is large enough. So we
remark that throughout this paper we fix the truncation
dimension of MPS $D=70$ and the numerics will not change with the
increase of the truncation dimension. Firstly, it is interesting
to notice that the fidelity is monotonously going down with the
increase of the parameter $\lambda$, implying that more
information of input qubits is missing with the increase of
$\lambda$. This is not surprising because we know the fidelity of
teleportation depends on quality of the quantum channel. When
$\lambda\rightarrow 0$ the ground states of
quantum Ising chain is
dominantly determined by the interaction and the
entanglement between the qubits of the channel is stronger,
while when $\lambda\rightarrow \infty$ the ground state is
dominantly determined by the second term of the Hamiltonian
in Eq.(\ref{eq:eps}) such that two qubits of the channel
 disentangled, leading to a vanishing fidelity
of teleportation.
%%%%%%%%%%%%%%%%%%%%%%%%%%
\begin{figure}
\center{
\includegraphics[scale=0.4]{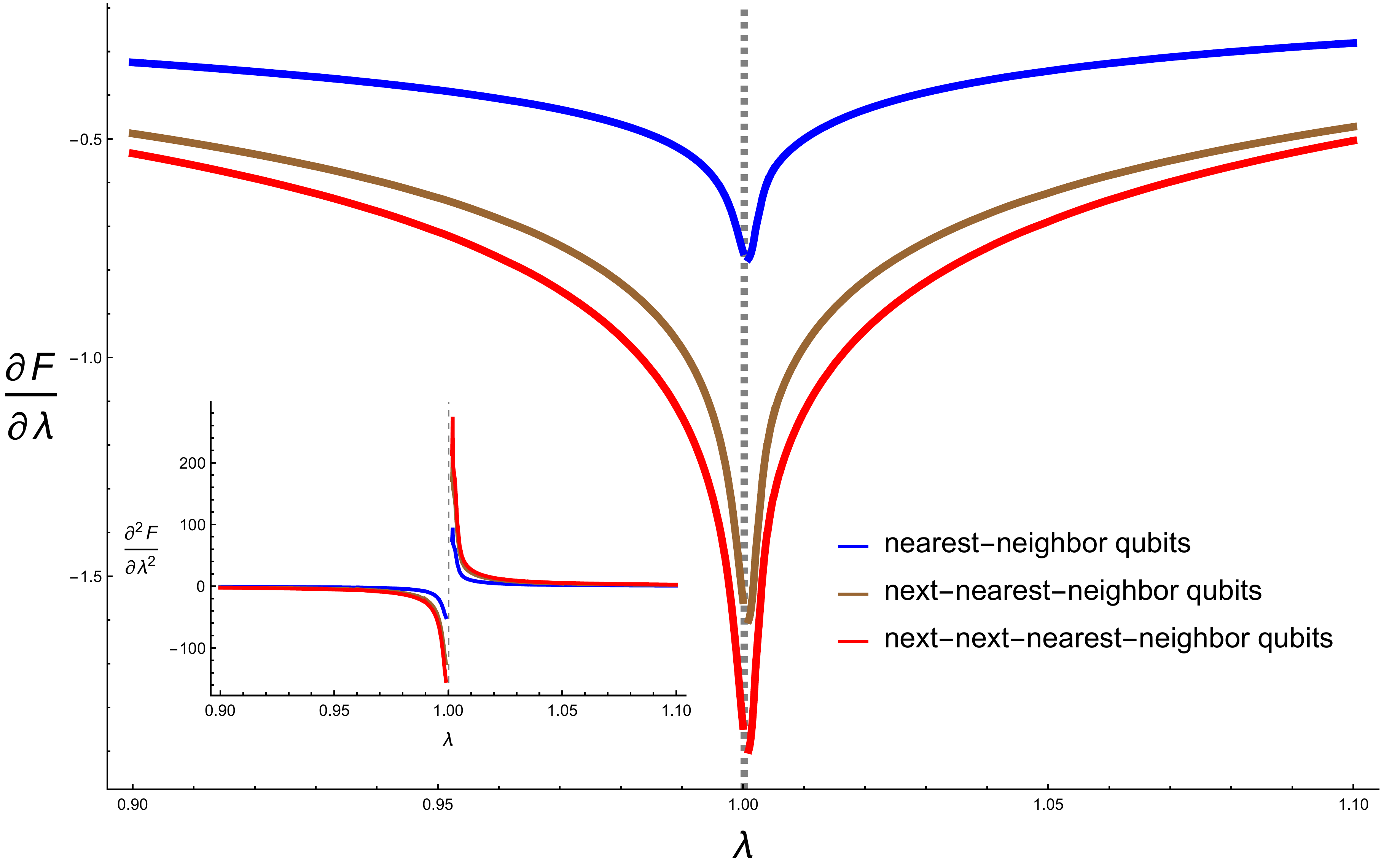} \\ \hspace{0.1cm}
\caption{\label{Fig5}The first order derivative of the fidelity $\frac{\partial F}{\partial \lambda }$ for the nearest-neighbor qubits, the next-nearest-neighbor qubits and the next-next-nearest-neighbor qubits with $\gamma=1$.}}
\end{figure}
%%%%%%%%%%%%%%%%%%%%%%%%%%%%%%%%

%%%%%%%%%%%%%%%%%%%%%%%%%%
\begin{figure}
\center{
\includegraphics[scale=0.4]{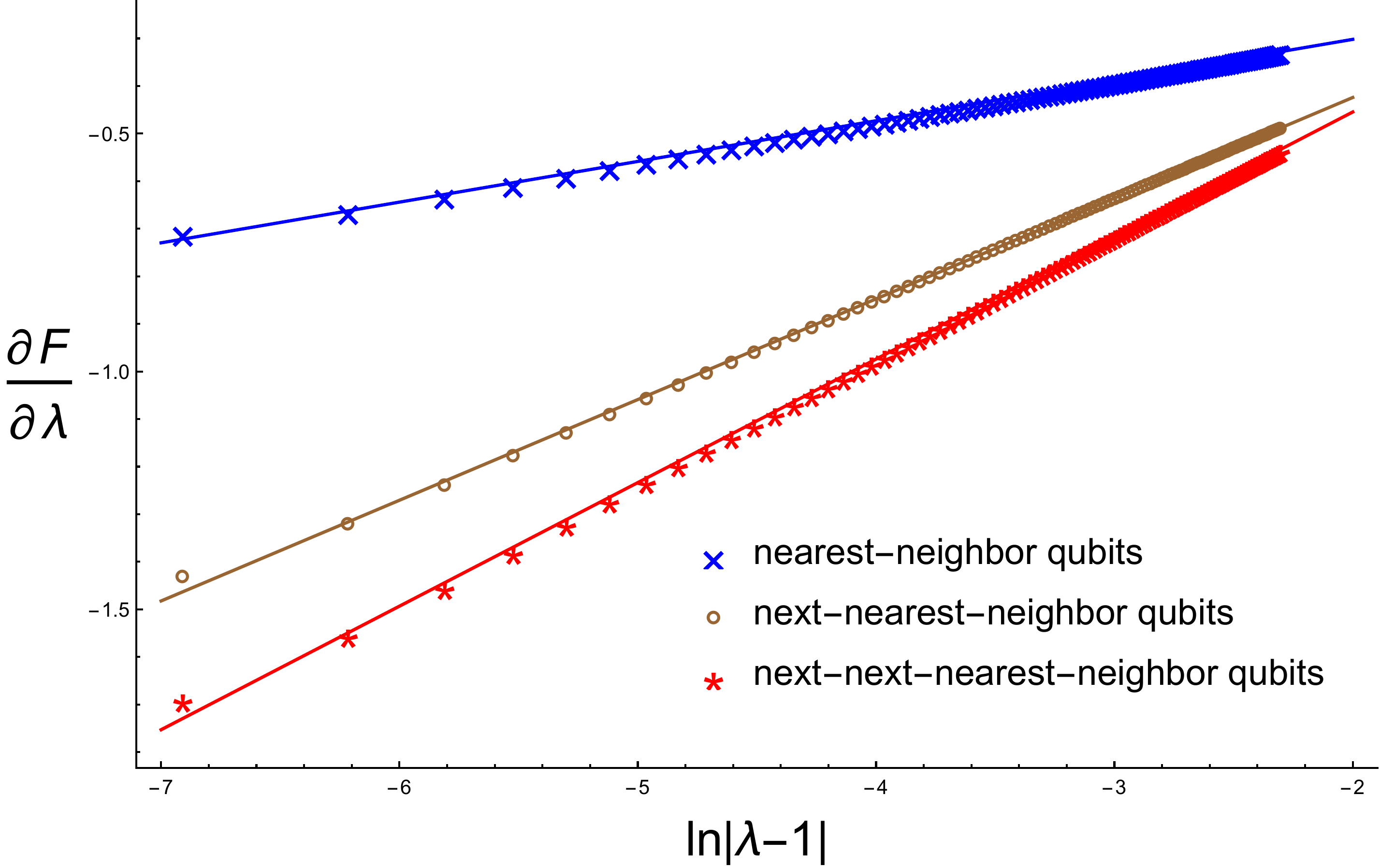} \\ \hspace{0.1cm}
\caption{\label{Fig9}The first order derivative of the
fidelity $\frac{\partial F}{\partial \lambda }$ is in direct
proportion to $ln\left | \lambda -1 \right |$  for the
nearest-neighbor qubits, the next-nearest-neighbor qubits and the
next-next-nearest-neighbor qubits with $\gamma=1$.}}
\end{figure}
%%%%%%%%%%%%%%%%%%%%%%%%%%%%%%%%

Next we notice that the fidelity goes down more quickly around the
critical point regardless of whether two qubits of the channel are
nearest neighboring or next-nearest neighboring or
next-next-nearest neighboring in the chain. To describe this
tendency more quantitatively we plot the first order derivative of
the fidelity $\frac{\partial F}{\partial \lambda }$ near the
critical point of the system in Fig.\ref{Fig5}. It is remarkable
to observe that this quantity exhibits a divergent behavior at the
critical point\blue{\footnote{One more evidence of this divergence is the the second-order derivative of the fidelity exhibits discontinuities at the critical point as shown in the insert of the Figure 7. }}, which is the main results obtained in this paper.
More precisely, we find that in critical region the first
order derivative of the fidelity $\frac{\partial F}{\partial
\lambda }$ is directly proportional to $ln\left | \lambda -1
\right |$, as illustrated in Fig.\ref{Fig9}. It implies that the
quantity $\frac{\partial F}{\partial \lambda }$ has a logarithmic
singularity at the critical point, which is just like the behavior
of the concurrence as firstly disclosed in
\cite{Osterloh:2002soe}.
One could notice that the first order derivative of the fidelity for the nearest-neighbor qubits should more diverged at the critical point from Fig.\ref{Fig5}. We clarify that this phenomenon are caused by the limitation of computer precision. And we compute the second derivative of the fidelity near the critical point as shown in inset of Fig.\ref{Fig5}, and we
find that it becomes discontinuous at the critical point.
In addition, from Fig.\ref{Fig9}, we could see
that when the value of $\lambda$ tends to $1$, i.e. $ln\left |
\lambda -1 \right |\rightarrow -\infty $, the deviation between
numerical values and the fitted lines becomes more evident.

Finally, we argue that this
phenomenon is general in the sense that it does not depend on the
parameter $\gamma$ in entangled Werner state. To show this we plot
the first order derivative of the fidelity for nearest-neighbor
qubits with different values of $\gamma$ in Fig.\ref{Fig6}.
%%%%%%%%%%%%%%%%%%%%%%%%%%
\begin{figure}
\center{
\includegraphics[scale=0.4]{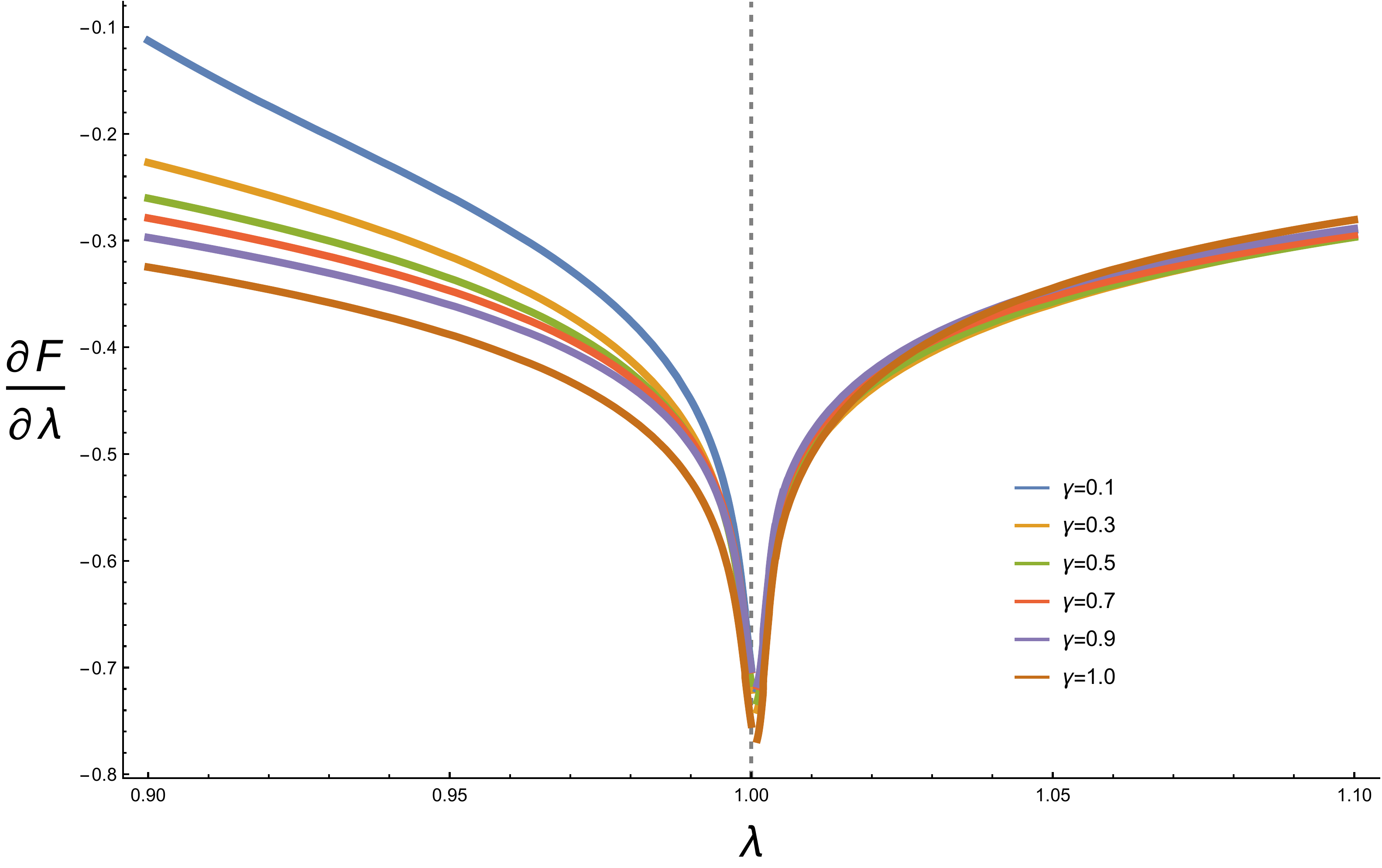} \hspace{0.4cm}
\caption{\label{Fig6}The first order derivative of the fidelity
$\frac{\partial F}{\partial \lambda }$ for nearest-neighbor qubits
with different values of $\gamma$.}}
\end{figure}
%%%%%%%%%%%%%%%%%%%%%%%%
The divergence of $\frac{\partial F}{\partial \lambda }$
can be understood from the perspective of entanglement.  Basically, we know the efficiency of
transmission through this quantum channel depends on two facets.
One is the entanglement between these two qubits made of the
channel, which may be called intrinsic entanglement; the other is
the entanglement between the channel and the environment, which is
composed of all other qubits in spin chain which have been traced
out. We intend to call this external entanglement. Obviously, the
stronger is the external entanglement, the lower is the
transmission efficiency. Furthermore, the degree of external
entanglement can be reflected by the entanglement entropy which is
defined as $S=-Tr(\rho_cln\rho_c)$. It is well known that this
quantity as
well as its derivative displays a peak
at the critical point, as
illustrated in Fig.\ref{Fig7}. Therefore, when $\lambda$ runs from
zero to infinity, as a global tendency the fidelity is largely
determined by the intrinsic entanglement and becomes smaller,
while near the critical point the system undergoes the most
prominent change with the parameter $\lambda$ and the quality of
the quantum channel is affected by this external entanglement with
environment most severely. As a reflection, the fidelity of
transmission goes down more quickly
at the critical point,
leading to the
divergence of $\frac{\partial F}{\partial
\lambda }$.

%%%%%%%%%%%%%%%%%%%%%
\begin{figure}
\center{
\includegraphics[scale=0.4]{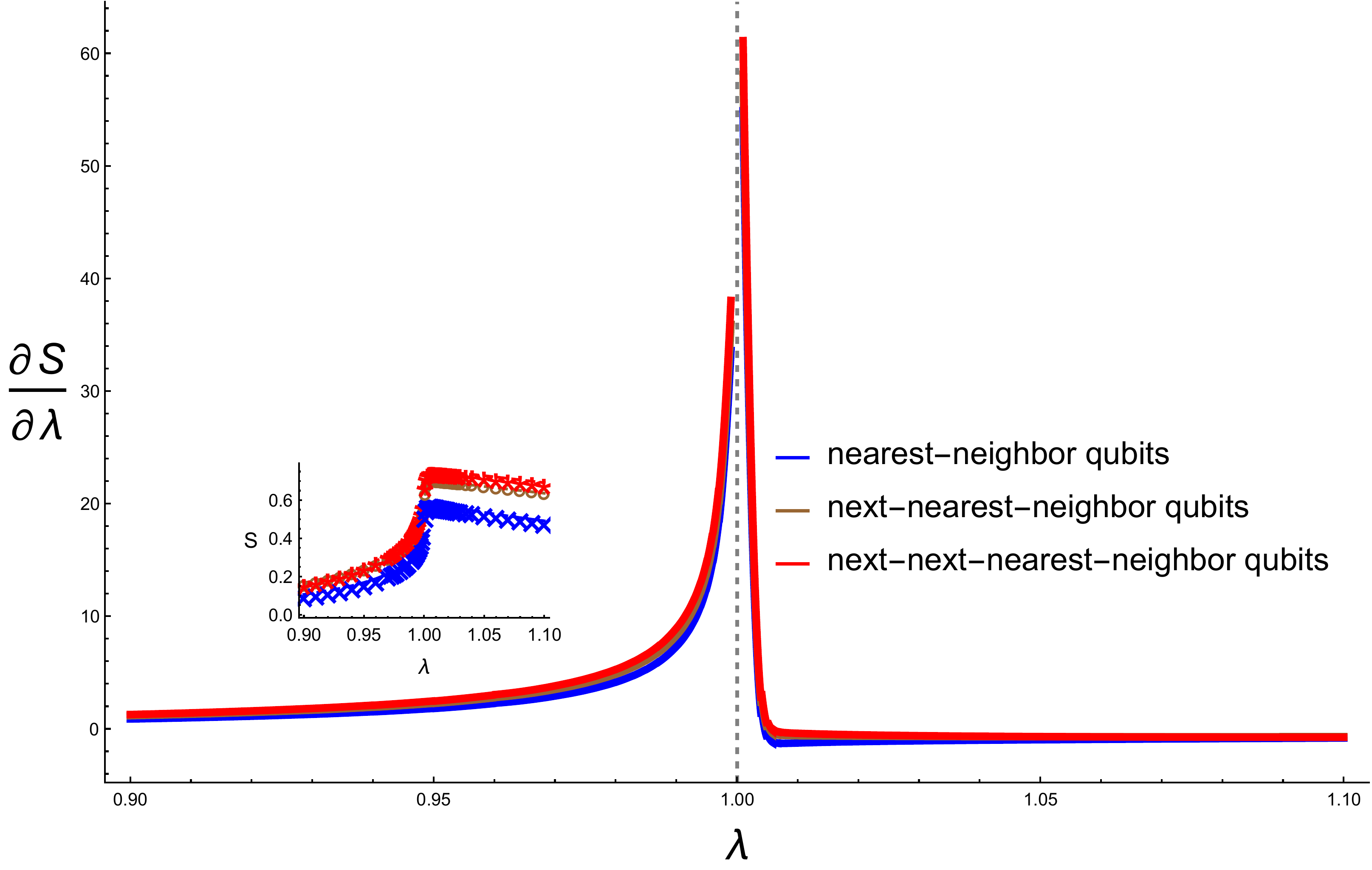} \\ \hspace{0.1cm}
\caption{\label{Fig7}The first order derivative of the external entanglement entropy
$\frac{\partial S}{\partial \lambda }$ for the nearest-neighbor qubits, the next-nearest-neighbor qubits and the next-next-nearest-neighbor .}}
\end{figure}
%%%%%%%%%%%%%%%%%%%%%%%%%%

\section{Conclusion and Discussion}
In this paper we have proposed a novel way to diagnose the quantum
phase transition by constructing a quantum channel with mixed
state for teleportation. In this circumstance we have found that
the first-order derivative of fidelity
exhibits a divergent behavior
at the critical point. Firstly, we intend to stress that what we
have observed may not be limited to the quantum Ising chain or the
Werner state considered in this paper. Instead, the relations
between the fidelity and quantum phase transition should be
general and the experiment on teleportation may also play a key
role in diagnosing quantum critical phenomenon. For instance, we
may consider another state called X-state \cite{Yu:2007} as input
state, whose form is given as $\rho _{X}=\frac{1}{4}\left ( \sigma
_{0}\otimes \sigma_{0}-\frac{2\gamma+1 }{3}\sum_{i=1}^{3}\sigma
_{i}\otimes \sigma _{i}+\frac{2\epsilon +1 }{3}\left ( \sigma
_{1}\otimes \sigma _{1}- \sigma _{2}\otimes \sigma _{2} \right )
\right )$.
 For simplify, we also find the extremal
behavior of $\frac{\partial F}{\partial \lambda }$ at the quantum
critical point, as demonstrated in Fig.\ref{fign10}. Secondly,
in contrast to all the previous references on the relation between
the fidelity and quantum phase transition, we have employed the
fidelity to measure the information loss during the transmission
rather than the Hilbert-Schmidt fidelity which only measures the
difference between two grounds states of a many body system. In
this sense, our paper has paved a new bridge linking condensed
matter physics to quantum information and may stimulate
experimentalists to explore more exciting phenomena in laboratory
such as the field of cold atoms.

%%%%%%%%%%%%%%%%%%%%%%%%
\begin{figure}
\center{
\includegraphics[scale=0.4]{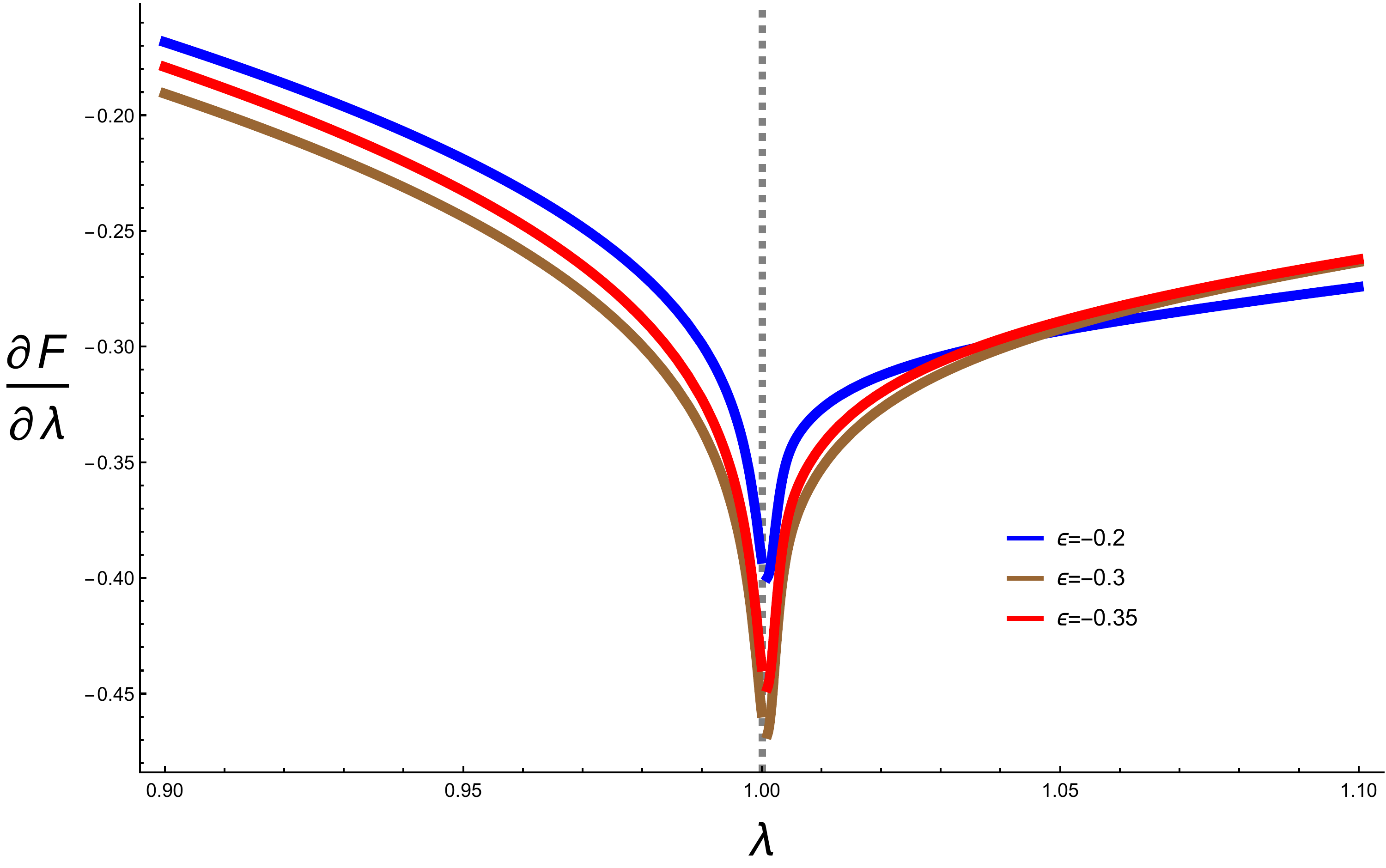} \hspace{0.4cm}
\caption{\label{fign10}The first order derivative of the fidelity $\frac{\partial F}{\partial \lambda }$ for the nearest-neighbor qubits with $\gamma=0.6$ for different values of $\epsilon$ by using X-state as input state (truncation dimension D =
70).}}
\end{figure}
%%%%%%%%%%%%%%%%%%%%%%%%

Finally, remarkable progress has been made in recent years on the
relation between condensed matter system and the geometry of
space-time, in which the entanglement plays an essential role in
describing the microscopic structure of space time and
understanding the emergent signature of
geometry\cite{Swingle:2009bg,Swingle:2012wq,Gan:2016vjt,Qi:2013caa,Czech:2015qta,Miyaji:2015fia}.
However, it was pointed out by Susskind
\cite{Susskind:2014rva,Susskind:2014moa} that entanglement maybe not enough and further quantum information quantities are
needed in our understanding of holography. One recent effort is a
conjectured duality\cite{takaya} between fidelity susceptibility
and the max volume of a codimension-one time slice in the Anti-de
Sitter. The system under question is a priori conformal invariant
at the critical point. This makes it very suitable
for investigating its gravity dual of the involved fidelity
within the framework of AdS/CFT correspondence. Moreover, the
feature of the fidelity that we have found in this simple model
may be helpful for us to investigate its role in the route of
reconstructing geometry by holography.

\begin{acknowledgments}
We are very grateful to Peng Liu, Zhuo-Yu Xian, Yi-Kang Xiao and
Xiang-Rong Zheng for helpful discussion. We also thank the
anonymous referees for helpful comments and suggestions. This
work is supported by the Natural Science Foundation of China under
Grant Nos. 11575195 (Y.L.) , 11465012(F.W.S.) and 11665016
. Y.L. also
acknowledges the support from Jiangxi young scientists (JingGang
Star) program and 555 talent project of Jiangxi Province.

\end{acknowledgments}

\end{document}